\documentclass[final]{./svjour2}
\usepackage{graphicx}
\usepackage{rotating}
\usepackage{amssymb}
\usepackage{mathptmx}
\usepackage[numbers]{natbib}
\makeatletter
\journalname{Journal of Low Temperature Physics}

\bibpunct{}{}{,}{s}{}{,}

\begin{document}

\title{Supersolid Phases of Cold Atom Assemblies}

\author{M. Boninsegni}

\institute{Department of Physics, University of Alberta,\\ Edmonton, Alberta, Canada, T6G 2G7\\
\email{m.boninsegni@ualberta.ca}}

\date{10.31.2011}

\maketitle


\begin{abstract}
We review recent theoretical results for soft-core Bose systems,   and describe the low-temperature supersolid ``droplet crystal" phase, predicted  for a broad class of soft-core interactions. We identify the conditions on the inter-particle interaction that render such intriguing phase possible, and outline  proposals for its observation. We argue this to be the prototypical supersolid, at least in the context of assemblies of ultracold atoms. 
\end{abstract}

\section{Introduction}
At the time of this writing, agreement has yet to be reached over whether the  Non-Classical Rotational Inertia (NCRI) in solid $^4$He at low 
temperature, first reported by Kim and Chan\cite{kc}, marks the first actual observation of the long sought supersolid phase of matter. The sheer 
weakness of the superfluid signal, the predicted resilience of superfluidity (SF) in the metastable, overpressurized liquid\cite{tie}, the ostensibly 
important role played by extended defects, as well as some puzzling experimental observations (such as the concomitant occurrence of the NCRI with stiffening of the crystal \cite{day}), have so far precluded a straightforward interpretation of the phenomenology in terms of any conventional framework of supersolidity. 
First principle numerical simulations of crystals of $^4$He, have yielded rather strong evidence that superfluidity does not take place in a perfect crystal\cite{sg,cc,galli}, and 
cast serious doubts on microscopic scenarios based  on Bose-Einstein Condensation (BEC) of  point defects, such as vacancies or interstitials \cite{pollet}.  
\\ \indent
While there exist dissenting views\cite{solito}, it seems fair to state that consensus is growing that the ground state of solid $^4$He is commensurate, i.e., free of point defects, and non-superfluid. Indeed, mounting experimental and theoretical evidence suggests that, if supersolidity occurs at all in helium, its  character is likely to deviate significantly from the physical picture laid out in the seminal works of Andreev and Lifshits\cite{Andreev69}, Chester\cite{Chester} and Leggett\cite{Leggett70}. In particular, extended defects such as grain boundaries\cite{pollet07} or dislocations\cite{pollet07b} might play a significant role in the phenomenology\cite{local}. 
\\ \indent
In order to make progress toward a more complete understanding of the physics of solid helium, and more generally of the supersolid phase, a useful first step seems to be the unequivocal identification of a simple  physical system, whose supersolid character can be established unambiguously,  by means of controlled measurements as well as reliable, first principle calculations.  The investigation of such a system may shed light on yet poorly understood aspects of the supersolid phase, which could in turn help  resolve the controversy over solid helium, as well as  facilitate the detection of supersolid behavior in other systems.
The question is, of course, which other physical system, besides helium, can be opined to feature a supersolid phase, in a region of its phase diagram accessible by current  technology. 
\\ \indent
Among  simple atomic or molecular condensed matter systems, helium has always been regarded as the ideal candidate to display supersolid behavior. Due to the very low reactivity at ordinary conditions of temperature and pressure, helium atoms can be to an excellent approximation 
considered elementary particles; for the most abundant isotope ($^4$He), these particles have spin zero,  and thus obey Bose statistics. On account of  its  low atomic mass, as well as of the weakness of the inter-atomic potential, liquid helium escapes crystallization at low temperature, under the pressure of its own vapor. Solidification occurs if moderate pressure is applied, but atomic excursions away from lattice sites remain significant in the crystalline phase. 
\\ \indent
The closest condensed matter system to helium, that may enjoy similar properties, is molecular hydrogen (H$_2$), also an assembly of Bose particles. Actually, the mass of a parahydrogen molecule is  half of that of a helium atom, which should lead to even greater quantum effects. However, the attractive well of the interaction between two such molecules is about three times deeper than that between two helium atoms. As a result, liquid hydrogen crystallizes at a relatively high (14 K) temperature, significantly above that at which BEC and SF are expected to occur. Albeit effects due to Bose statistics are detectable in the momentum distribution of liquid parahydrogen near freezing\cite{bon09}, in  general the behavior of condensed molecular hydrogen is much closer to that of a classical system than to helium.  Indeed, while some theoretical\cite{mezza1,mezza2} and experimental\cite{toennies} evidence of superfluidity (and even possible {\it supersolidity}\cite{mezza3}) in small clusters of parahydrogen (comprising around twenty molecules) has been reported, agreement is virtually unanimous that bulk condensed parahydrogen displays no superfluid behaviour\cite{bon04,clarkh2}.
\\ \indent
As it turns out, the physics predicted for solid $^4$He (or, parahydrogen), as it emerges from first-principle quantum simulations, is essentially a direct consequence of the strong repulsive core of the interatomic potential
at short distance (less than $\sim$ 2 \AA). It is the repulsive core that determines most of the thermodynamic properties of  helium and other quantum solids and liquids, witness the fact that  a very simple model of Bose hard spheres reproduces rather accurately the phase diagram of condensed helium \cite{huang}.
\\ \indent
Computer simulation studies of classical crystals, based on the Lennard-Jones potential (also featuring a hard core), suggest that, much like  in a
 highly quantal solid like helium\cite{pollet}, a uniform gas of point defects (vacancies or interstitials) is unstable against phase separation \cite{ma}.  This suggests that the basic physics of the crystals is dominated by the strong interaction among particles, and that quantum effects cannot really alter it qualitatively (unlike in the liquid phase).  One is therefore naturally led to pose the following questions: can a supersolid phase of matter be underlain, in a many-body system, by a pair-wise interaction among elementary constituents featuring a  ``soft" repulsive core at short distances  ? And, if that is the case, do such physical systems exist in nature, or are they at least artificially realizable ? 
\section{A different scenario: modified interactions and ultracold gases}
If one wishes to retain the simple picture of ``elementary" particles interacting via a static pair potential \cite{note,note2},  spatially confined ultracold gases provide an alternative, in many respects more promising option to the controlled observation of the supersolid phase. 
It is now seventeen years since the first successful observation of BEC in Rubidium gas, cooled down to temperature in the nanoKelvin range \cite{tutti,pethick}.  Impetuous  scientific and technological progress has rendered   the field of ultracold atoms  the  playground of choice, to which one can turn to address outstanding questions  in condensed matter and many-body physics.\\ \indent
Dilute assemblies of cold atoms represent almost a textbook many-body system,  {\it a}) virtually free from the imperfections and ``background noise" that often mask subtle physical effects in a sample of condensed matter, and  {\it b}) affording a degree of control that  no naturally occurring solid or liquid can match.  
Last but not least, there exist a number of techniques whereby the interaction between atoms or molecules can be altered, making it virtually an adjustable parameter. The simplest example of such techniques, is  the so-called {\it Feshbach resonance} \cite{pethick}, allowing  one to vary the strength of the (short-ranged) interaction between two atoms or molecules, and even  reverse its sign (i.e., turned from repulsive to attractive, or vice versa).
But then, if novel artificial inter-particle potentials can be created, the field is wide open for the search for a  specific type of interactions, or broad 
class thereof, for which supersolid (or other yet unexplored)  phases may occur.

\subsection{Dipolar systems}
As mentioned above, the repulsive core of virtually {\it any} known  interaction among atoms or molecules,  is what prevents microscopic scenarios of supersolidity based on point defects from occurring, as such defects have the tendency to cluster together. On the other hand, there exist theoretical predictions of supersolid behavior in two-dimensional systems wherein pair-wise interaction feature a slow decay at large inter-particle separations, e.g., $\sim$ $1/r^n$, with $n \le 3$. 
\\ The physical picture is radically different, in this case, from that based on BEC of point defects, and it is based on the  peculiar behavior of a purely repulsive dipolar system in the proximity of crystallization. In an ordinary first-order phase transition, there exists a finite range of density $(\rho, \rho+\delta\rho)$ wherein no homogeneous phase is thermodynamically stable. Rather, two phases of different density (i.e., $\rho$ and $\rho+\delta\rho$), coexist, separated by a macroscopic interface.  It can be shown, however, that in the presence of long-ranged interactions (such as the dipolar) the energy associated to such an interface contains a negative term, diverging logarithmically in the thermodynamic limit \cite{Spivak}.
As a result,  ordinary coexistence is energetically disfavoured. On approaching the transition  from the low-density (e.g., the liquid) phase, 
the system may lower its free energy by embedding sufficiently large solid domains (i.e., macroscopic
``clusters") inside the liquid. In the low temperature limit, two effects are predicted:  the transition of the liquid to a superfluid, and the crystallization of solid bubbles into a  lattice superstructure, resulting
in a  global supersolid phase (in fact, a whole set of
different such phases \cite{Spivak}).

A well-defined way of testing the above prediction in ultracold gases, makes use of atoms or molecules possessing a finite electric dipole moment. These particles can be confined to quasi two dimensions,  by means of an external harmonic potential in the direction perpendicular to the motion (the so-called ``pancake" geometry). Upon aligning all dipoles in the direction perpendicular to the plane, by means of a strong external electric field, one can study a system of Bosons interacting  via a purely repulsive potential \cite{minkia} of the form $1/r^3$.
\\ \indent
There is a subtle aspect in the derivation of the above predictions, however. 
For,  the negative contribution to the surface tension is proportional
to $(\delta \rho)^2 \ln (R \rho^{1/2})$,
where $R$ is the droplet size. If the jump in density $\delta \rho$ at the
phase transition is small (and quantum Monte Carlo simulations\cite{minkia,cagate,cool} suggest that this 
may indeed be the case), the characteristic value of $R$ required to observe the above
scenario may be exceedingly large,
for practical purposes outside the reach of realistic experimental or even numerical setups.
\\ \indent
The  remainder of  this article  focuses on a different  class of interactions,  featuring a  soft-core potential at short inter-particle distance,  one which does not grow arbitrarily, but plateaus to a finite value. 
Below, we  review  theoretical predictions, based on first principle numerical simulations, of a system of  such soft-core bosons, in two dimensions. The purpose is twofold: on the one hand,  such a system can be shown to support a supersolid phase, whose physics turns out to be quite intuitive, arguably simpler than the scenarios that have been thus far considered for solid helium. Secondly, because this type of interaction appears to be realizable in assemblies of cold atoms, by means of  the so-called Rydberg blockade, cold atoms may provide an entirely new, likely more direct pathway to the observation in the laboratory of a supersolid phase.
\subsection{Soft-core bosons}
\begin{figure}[tbp]
\includegraphics[width=3.0in]{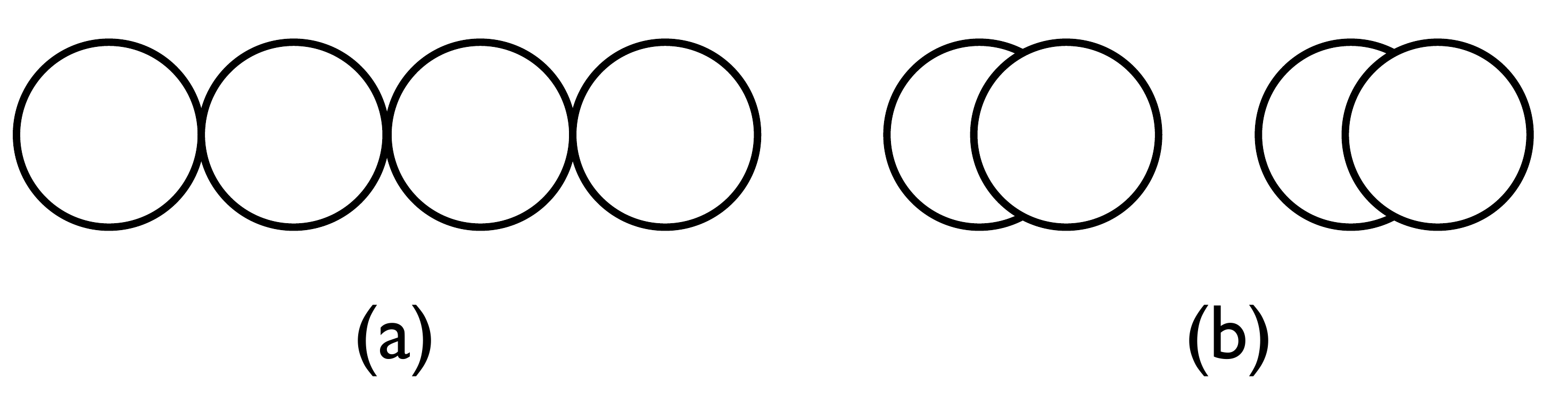}
\caption{One-dimensional system of particles interacting via a soft-sphere potential, equal to $V > 0$ if the distance between two particles is less than their effective diameter $a$, and zero if it is greater. As the distance between nearest neighboring particles approaches  $a$,  arrangement shown in (b) is favored over that shown in (a).
}
\label{fm1}
\end{figure}

In order to understand how a pair-wise interaction featuring a soft repulsive core at short inter-particle separation can lead to the appearance of a supersolid phase, it is useful to elucidate first the nature of the classical crystalline phase that occurs in a system characterized by this kind of interaction. To this aim, we consider the simplest soft-core pair  potential,  i.e., one that is equal to some energy $V>0$ if particles are less than a distance $a$ apart, zero otherwise. \\ Imagine lining up particles  interacting via this potential  in one dimension, as shown in Fig. \ref{fm1}.
As the density of the system approaches the point where each particle barely starts touching its two nearest neighbors, as shown in Fig. \ref{fm1}(a), the potential energy per particle discontinuously goes from zero to $V$, in correspondence to a classical crystal with lattice constant $a$. However, the system can lower its energy by one half, by allowing each other particle to overlap in part with one of its  neighbor (on the left, for example), i.e., effectively doubling the lattice constant of the crystal, as each unit cell contains now two particles. On increasing the density, the number of particles per unit cell increases, while the lattice constant remains unchanged. Crucial to this effect, of course, is the finite energy ``cost" associated to particle overlap.
\\ \indent
The subject of {\it multiple occupation crystals} (or ``cluster crystals'',  as they are often referred to), has been extensively investigated in classical materials science, where they arise in the context of colloids and macromolecules \cite{likos,liu, archer,mladek}. In particular, even though the qualitative argument offered above made use of the simple barrier potential, a quantitative criterion has been formulated, that allows one to predict the average number of particles $K$ in the unit cell of the cluster crystal, based on general features of the soft-core interaction\cite{likos}.\\ \indent
How does quantum mechanics alter this simple picture ? In two fundamental ways: {\it a}) owing to quantum delocalization, particles can ``hop" to adjacent clusters, and {\it b}) an effective attraction among identical particles sets in in the case of Bose statistics, as  exchanges of identical particles results in a lower  kinetic energy (this is, of course,  the same physics that gives rise to Bose-Einstein Condensation).

Consider for definiteness a system of identical Bose particles of mass $m$, and let their interaction be described by a pair-wise potential, only depending on the relative distance of two particles, for which an effective range $a$ can be defined. If all distances are expressed in units of $a$, and $\epsilon_\circ\equiv \hbar^2/ma^2$ is the energy unit, the (dimensionless) many-body Hamiltonian can be expressed as follows:
\begin{equation}
\hat H = -\frac{1}{2}\sum_i\nabla^2_i + \sum_{i < j} v(r_{ij})
\end{equation}
where $r_{ij}$ is the distance between $i$th and $j$th particles. The only requirement on $v(r)$, is that $v(r\to 0) = v_\circ \sim 1/r_s^2$, where $r_s$ is the (dimensionless) mean inter-particle distance. In other words, the  energy barrier associated to the repulsive core of the interaction, should not be much greater than the characteristic kinetic energy\cite{notey}. \\ \indent
In order to render the discussion more quantitative, we  now focus on the phase diagram of a specific model of soft-core Bosons. We assume the simplest form of the pair potential, namely the ``barrier" one discussed above
\begin{equation}\label{pippo}
V(r) = D \ \Theta(r-1)
\end{equation}
where $\Theta$ is the Heaviside's function and the potential strength $D$ is expressed in units of $\epsilon_\circ$. As it tuns out, this simple model embodies all the physics of interest that we wish to present here. We postpone therefore the discussion of more more realistic interactions, until after establishing the main physical results for the barrier potential. Henceforth, we shall assume that the system is confined in two dimensions.
\begin{figure}[tbp]
\includegraphics[ width=3.0in]{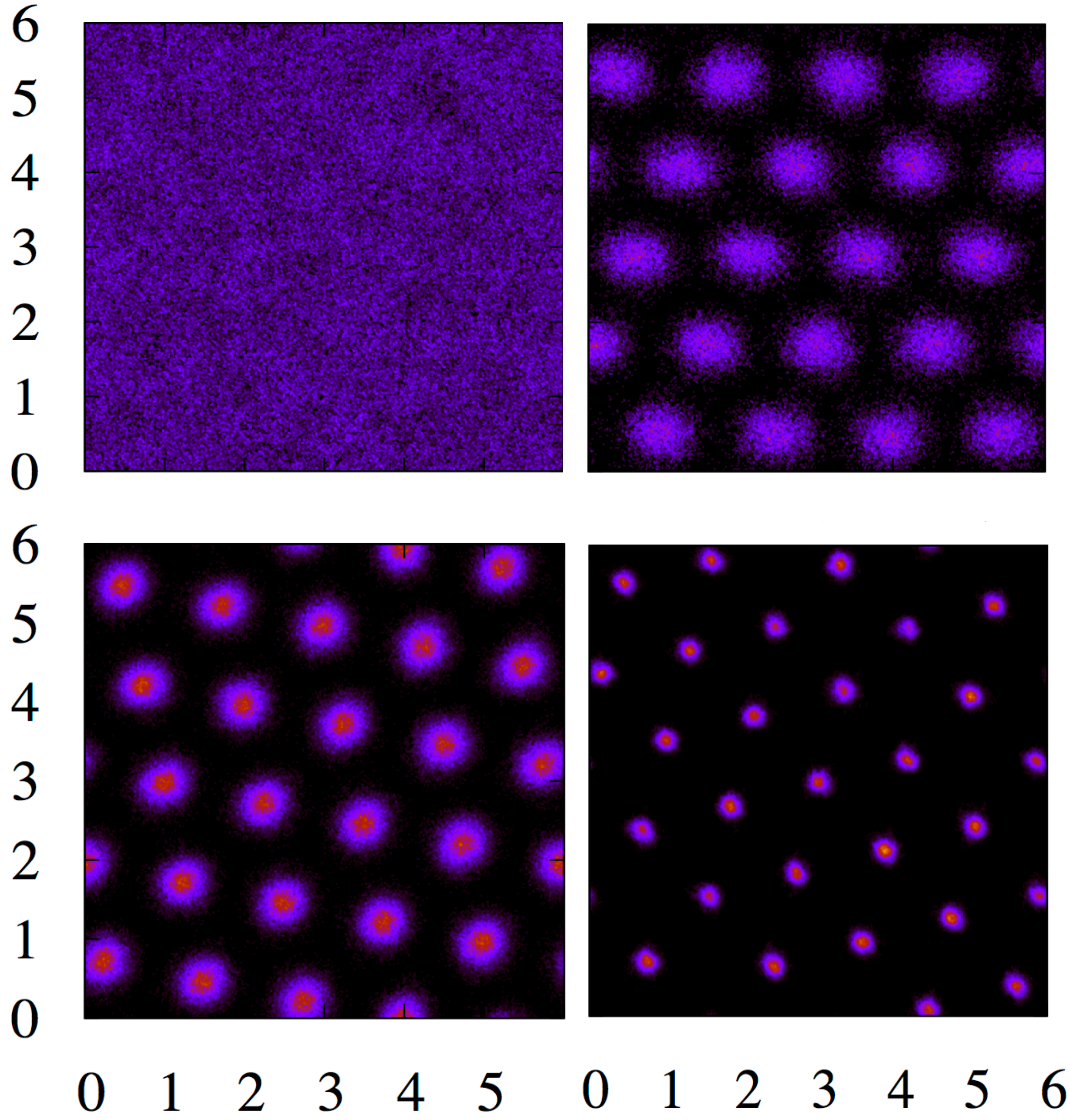}
\caption{{\it Color online}. Density of particles for a system of soft-core bosons in two dimensions, at low temperature. The nominal interparticle distance is $r_s=0.5$ in all four cases shown. The strength of the interparticle potential is $D$=1 (top left panel), $D=5$ (top right), 
$D$=10 (bottom left) and $D$=100 (bottom right). The number of particles in each simulated system is $N$=144. All lengths are expressed in units of the soft-core radius.
}
\label{fm2}
\end{figure}
\\ \indent
The thermodynamic properties of a two-dimensional system of spin-zero bosons, interacting via a soft-disk pair potential, can be investigated by means of computer simulations, which have the advantage of being essentially {\it exact}, uncertainties being reducible in principle to arbitrary degree, given enough computer time. In particular, computer simulations based on the Worm Algorithm \cite{worm, worm2} have emerged in recent years as a powerful methodology, affording both  numerical accuracy as well as physical insight. It is particularly well suited to investigate Bose  superfluids, in that it affords the simultaneous, direct computation of the superfluid fraction, using the well-known {\it winding number} estimator of Pollock and Ceperley \cite{pollock}, as well as of  the one-particle density matrix. 
\\ \indent
The phase diagram of a two-dimensional system of bosons interacting via the potential (\ref{pippo}) has been recently investigated by Saccani {\it et al.}\cite{saccani}, by means of computer simulations making use of the continuous-space Worm Algorithm. We briefly review the basic physical results here.
\\ \indent
Fig. \ref{fm2} shows the density of particles in the low temperature limit (typically, results become temperature-independent for  $T\lesssim \ 
\epsilon_\circ$, i.e., they are representative of ground state physics), computed for a two-dimensional system of soft disks. The nominal inter-particle distance is $r_s=0.5$ is all cases shown; on the other hand, the strength $D$ of the interaction (\ref{pippo})   increases hundredfold from top left ($D=1$) to bottom right ($D=100$). For weak interactions, the system is in a gas phase, the density being uniform (top left panel of Fig. \ref{fm2}), but the formation of a solid crystal of clusters of particles,  is clearly seen for $D \gtrsim 2$. The number of particles $K$ per cluster remains remarkably constant (around 7 particles per clusters) as $D$ is  raised from a value of 5 (top right panel in Fig. \ref{fm2}) to 10 (bottom left), to 100 (bottom right).
\\ \indent
The spatial extension of clusters decreases as $D >> 1$, i.e., they become more and more compressed. In that limit, particles in the same cluster tend to ``pile up", in   order to limit their overlap with other particles, and its associated energy cost,  to those particles in the same cluster. 
Particle hopping to adjacent lattice sites, is a process that involves tunnelling under a potential barrier $\sim KD$, while the distance between nearest-neighboring clusters is $\sim\sqrt K$.  In a range of values of $D$ (see below), this leads at low $T$ to a superfluid transition, as particle hopping establishes phase coherence throughout the system. 
\begin{figure}[tbp]
\includegraphics[width=3.6in]{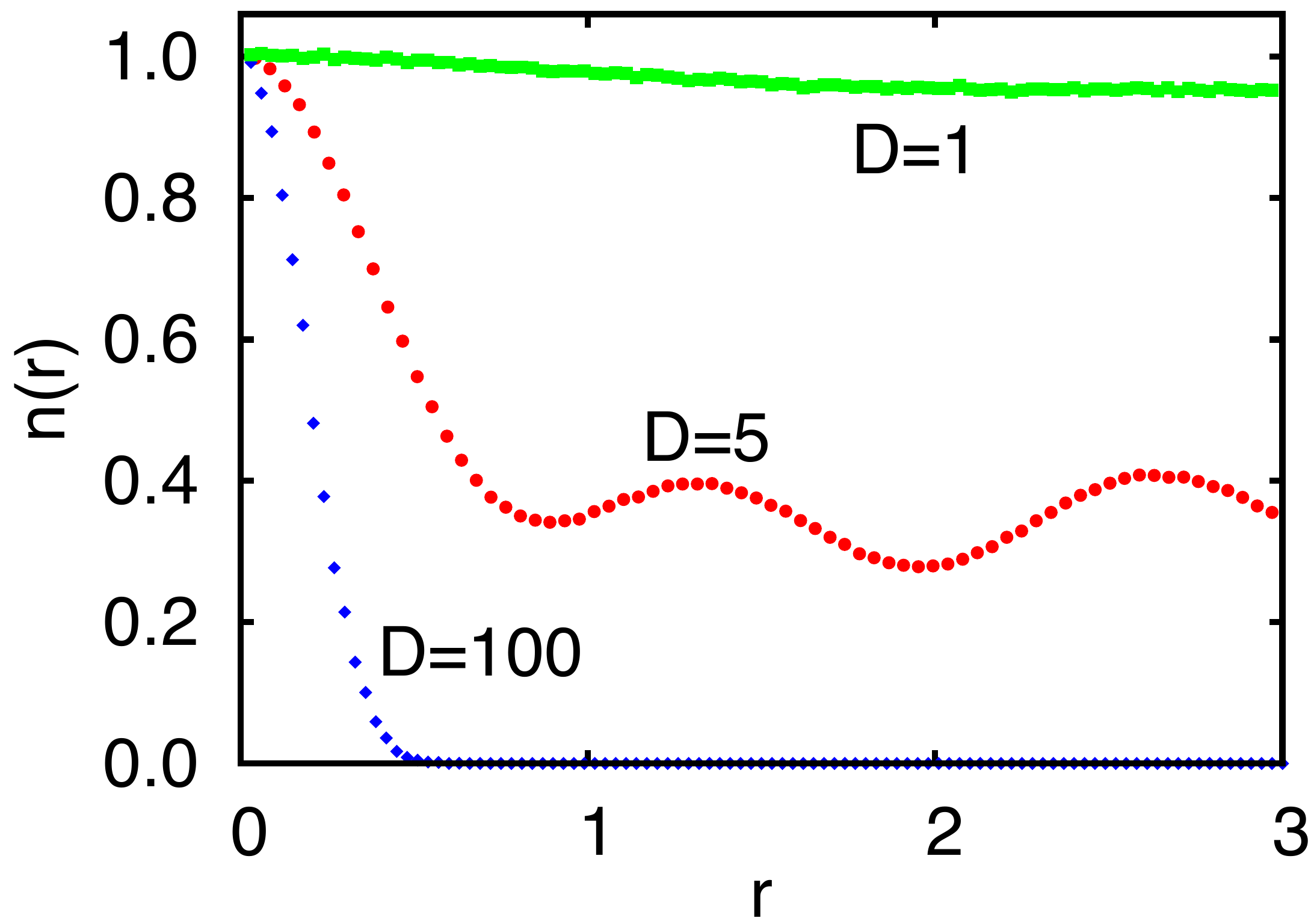}
\caption{{\it Color online}. One-body density matrix $n(r)$ for a system of soft disks, at different values of the barrier strength $D$ (corresponding to three of the four shown in Fig. \ref{fm2}). The temperature is taken sufficiently low that no significant change is observed on lowering it even further, within the statistical  uncertainties of the calculation, i.e., these are essentailly ground state results. The 
nominal interparticle distance is $r_s=0.5$ in all  cases shown.  All lengths are expressed in units of the soft-core radius.
}
\label{fm3}
\end{figure}
\begin{figure}[h]
\includegraphics[width=3.0in]{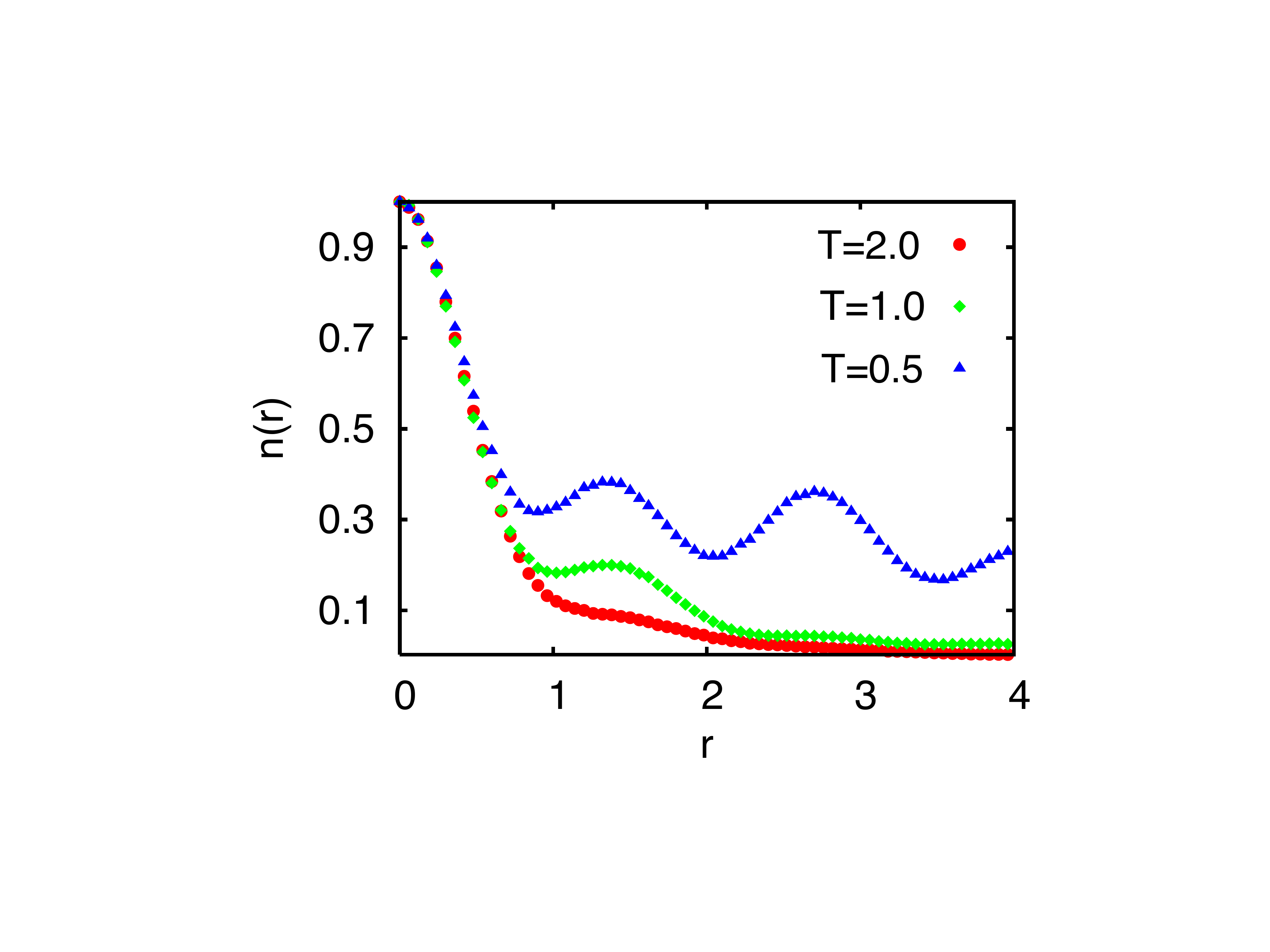}
\caption{{\it Color online}. One-body density matrix $n(r)$ for a system of soft disks, computed at different temperatures (in units of $\epsilon_\circ$), for a value of the barrier strength $D$=5.  
The  nominal interparticle distance is $r_s=0.5$ in all  cases shown.  All lengths are expressed in units of the soft-core radius. 
}
\label{fm4}
\end{figure}
\\ \indent
This is quantitatively shown in Fig. \ref{fm3}, which displays  low-temperature results for the 
circularly and translationally averaged one-body density matrix, defined as
\begin{equation}
n(r)=\frac{1}{4\pi \ {\cal V}}\  \int d\Omega\ \int d^3r^\prime\ \langle\hat\psi^\dagger({\bf r^\prime}+{\bf r})\hat\psi({\bf r^\prime})\rangle
\end{equation}
where ${\cal V}$ is the volume of the simulation cell, $\int d\Omega$ represents angular integration and $\langle ... \rangle$ stands for thermal average. Three different physical regimes can be identified. At low values of $D$, the system is a uniform gas ($D=1$), and consistently $n(r)$ displays a slow power-law decay. At the opposite end, namely for strong inter-particle interactions, the system is an insulating droplet crystal; thus, $n(r)$ decays rapidly, falling to essentially unmeasurable values  for $r \gtrsim r_s$, i.e., outside the nominal inter-particle separation. This is because tunnelling between adjacent particles is exponentially suppressed by the height of the effective potential barrier seen by an individual particles traveling across droplets.
\\ \indent
For intermediate couplings, (i.e., $D\sim 5$), the one-body density matrix mirrors the dual character of the system. On the one hand, $n(r)$ decays slowly at long distances, in conformity with the presence of a superfluid regime described by Kosterlitz-Thouless theory\cite{kt}. 
Concurrently, it displays  marked oscillations, reflecting the underlying crystalline arrangement of the clusters. These oscillations arise microscopically from tunnelling of particles across adjacent clusters, and give rise to satellite peaks  in the Fourier transform of $n(r)$, namely the momentum distribution, which in cold atom assemblies lends itself to experimental imaging  by means of time-of-flight measurements\cite{bloch}. Thus, the supersolid character of the system could be experimentally ascertained in a rather straightforward way. The emergence of off-diagonal long-range order, coupled with the solidlike oscillations, is clearly shown by the temperature dependence of the $n(r)$, displayed in Fig. \ref{fm4}.
\begin{figure}[h]
\centering
\includegraphics[ width=3.5in]{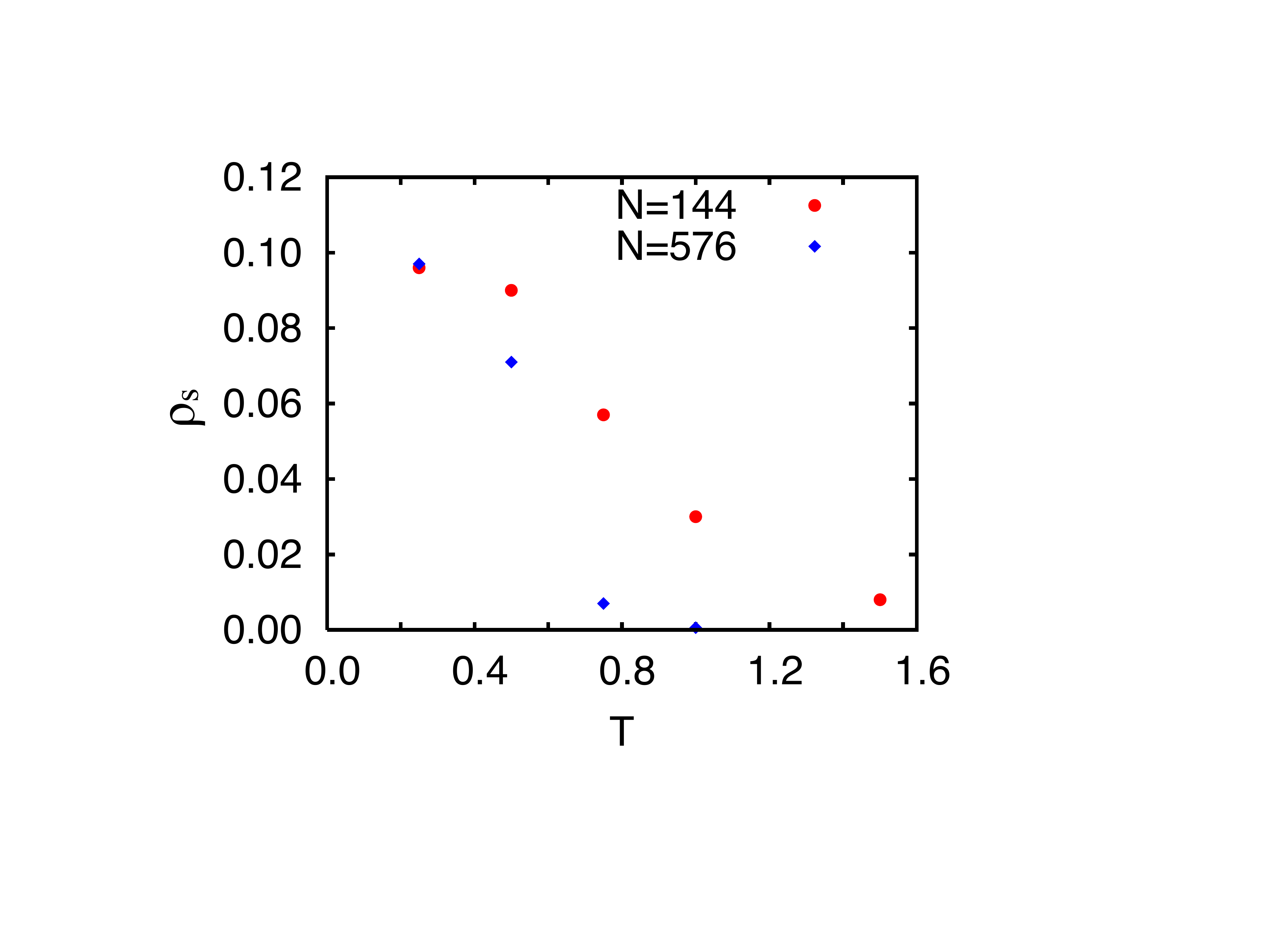}
\caption{{\it Color online}. Superfluid fraction $\rho_s$, computed as a function of the temperature (expressed in units of $\epsilon_\circ$) for simulated systems comprising $N=144$ (circles) and $N$=576 (diamonds) particles. The nominal inter-particle distance is $r_s=0.5$, whereas the value of the coupling $D$=5. Statistical errors are smaller than the sizes of the symbols.}
\label{fm5}
\end{figure}
The superfluid character of the droplet crystal phase can be established by direct computation of the superfluid density\cite{saccani}. A typical result is shown in Fig. \ref{fm5}, for two simulated systems comprising different numbers of particles. As in any numerical study based on computations performed for systems of finite size, results must be obtained for several systems, comprising different numbers of particles, in order to carry out extrapolation of the physical estimates to the thermodynamic limit. The results shown in Fig. \ref{fm5} are consistent with a superfluid transition in the Kosterlitz-Thouless universality class \cite{kt}. It is noteworthy that the superfluid fraction does not saturate to a value of 100\% in the $T\to 0$ limit, consistently with the prediction of Leggett\cite{Leggett70} for a superfluid system breaking translational invariance. 
\\ \indent
Interestingly, it is possible to ascertain that, even in the insulating (e.g., non-superfluid) droplet crystal phase, individual droplets are superfluid, as shown by the statistics of permutation exchange cycles\cite {saccani}. However, the transition of the system to a globally superfluid phase can only be established numerically in a well-defined range of values of $D$, as tunnelling is exponentially depressed  with increased $D$. Because the computational methodology adopted here is a {\it finite-temperature} one, we cannot make  a definitive prediction of  a {\it quantum phase transition} at $T$=0, driven by the parameter $D$, between a supersolid and an insulating droplet crystal, based on the results of the simulation alone; however, our results are certainly consistent with that conclusion.
\\ \indent
The physical behavior of the supersolid droplet crystal phase is certainly reminiscent of that of a Josephson junction array.
Futhermore, for  a system of identical bosons, one is naturally led to think of this many-body system in tems of a lattice Bose Hubbard model (BHM), which is known to display a superfluid ground state at low temperature\cite{fisher}.  It is important to note, however, that in this case the lattice is {\it self-assembled}, i.e., not imposed mathematically through  space discretization, or experimentally (e.g., by means of optical lattices). It is this crucial difference, that ultimately allows one to speak meaningfully of a {\it supersolid} phase in this context\cite{cazzate}.
\subsection{Rydberg blockade}
The physics illustrated above would reain of merely ``academic" interest, were it not because a well-defined procedure exists to realize a soft-sphere type interaction in an assembly of ultracold atoms, by means of a procedure known as {\it Rydberg blockade} (RB).  Such a physical mechanism was initially introduced as a device to manipulate quantum information stored in collective states of mesoscopic ensembles \cite{lukin}, but it also has  been recently proposed as a way to engineer a novel type of interaction potential  between cold atoms. Specifically, the modified interaction ``flattens off", and  remains essentially constant below some characteristic ``cut-off" distance $a$ \cite{henkel10}. 
\\ \indent
The soft sphere model should capture the main features of the interaction potential arising in the RB regime. However, it is necessary to assess the robustness of the main physical predictions illustrated above,  by studying systems characterized by more realistic model interactions, closer to those which could be realized in practice. In particular, one need assess any possible dependence on the physics of interest on the detailed features of the potential near the cut-off distance, as well as any  influence of a long-range tail of the interaction itself.
\\ \indent
As it turns out, the main physical results, chiefly the presence of a supersolid phase, can be shown to be largely independent on details of the interaction, beyond the presence of a soft core. Thus, the droplet crystal phase ought be observable  under relatively broad experimental conditions. 
For example, in the original work  in which the droplet crystal phase was first predicted\cite{cinti}, by means of computer simulations  like the ones described above, it was found that the same physics takes place in a two-dimensional system of bosons interacting through the following potential:
\begin{equation}\label {pipo}
V(r) \propto \frac{1}{r^3+a^3}
\end{equation} 
which also has a soft core at short distance, but a smooth behavior near the characteristic cut-off distance $a$, as well as a dipole-like long-range tail. The  potential (\ref{pipo}) provides a realistic description of the specific type of interaction between pair of cold atoms that could be realized in the laboratory, via the RB \cite{pup}. 
All of this constitutes  ``circumstantial" evidence that the presence of a ``plateau" at sufficiently short distances, namely a range within which the potential energy of interaction is either constant or varies very slowly as two particles are brought  closer together, is the crucial feature that enables the formation of the droplet crystal phase, which turns supersolid at low temperature. Indeed, recent studies of two-dimensional Bose systems in which particles interact via a repulsive Yukawa potential  show that no droplet crystal  (nor any other supersolid phases) exist in that system\cite{reatto}. The Yukawa potential has  an  infinitely strong repulsive core in the $r\to 0$ limit, which, despite its very slow growth, suffices to prevent the occurrence of the physics discussed here. Indeed, the basic physics of solids made of Yukawa bosons is the same as that of solid $^4$He. It is also worth mentioning that no supersolid phase has also been thus far observed in numerical studies of purely repulsive dipolar systems in two dimensions.
\\ \indent
It has been recently conjectured\cite{reatto}, that a {\it necessary} condition for the occurrence of a supersolid phase (presumably of the ``droplet crystal" kind), may be that the Fourier transform of the two-body potential go negative (i.e., that the potential become attractive) in some range of $k$. 
Such a condition is {\it not sufficient },  as the purely repulsive soft-core interaction, whose Fourier transform is oscillating and indeed goes negative for some values of $k$, does {\it not}    allow for a supersolid phase if the repulsive potential barrier is too high\cite{saccani}. 
\\ \indent
Now, obviously any realistic interaction displays a {\it hard} core at sufficiently short distances, as electronic clouds of different atoms are prevented by Pauli exclusion principle from  spatially overlapping. Therefore, the relevance of any soft-core model ultimately hinges on a substantial difference in scales between the characteristic radius of the region in which the repulsive part of the potential softens, and the (much smaller) radius of the hard core. Because the softening of the potential at short distances is an artifical feature, arising from the RB, it seems conceivable that one may be able to tune the parameters in order to realize such a condition.
\\ \indent
On the other hand, the behavior of the potential at large inter-particle separation is irrelevant to the existence of the supersolid phase described above, which has been predicted in theoretical studies in which different power law decays of the long-range tail were considered\cite{henkel10,cinti}.
\section{Conclusions}
After decades of intense investigation, aimed at identifying a supersolid phase in solid helium, it now appears as if such a phase might be detected and explored more easily in a different context, namely in cold atom assemblies. Besides providing  the experimenter with a remarkably ``clean" and controllable many-body system, the existence of techniques capable of fashioning the desired inter-particle interactions pave the way to the observation of novel phases of matter, difficult to achieve or non-existent in ordinary condensed matter physics.
\\ \indent
In particular, a physical mchanism known as the Rydberg blockade can give rise to pair-wise potentials between cold atoms, featuring a relatively ``flat" region at short inter-particle separations. As shown above, this spefici property can underlie a supersolid phase, consisting of a self-assembled crystal of superfluid droplets. Tunnelling of particles among adjacent droplets can establish phase coeherence and give rise to a supersolid phase, whose most direct experimental signature are secondary peaks in the momentum distibution. 
\\ \indent
Computer simulations give evidence that such phase should be observable under relatively broad conditions, as it is insensitive to the detailed behavior of the inter-particle potential at long distances. The same physical behavior is also observed with different potentials featuring a flat region at short distance. Hard core potentials, on the other hand, do not lead to a supersolid phase, even those whose growth at short distance is slow, like the Yukawa potential. 
\\ \indent
To our knowledge, no condensed matter system (either experimentally known or even toy model) has been predicted to display the supersolid phase illustrated here. The remarkable intuitiveness and simplicity of the supersolid droplet crystal phase makes it not only a promising candidate for a direct, unambiguous observation of the supersolid phenomenon, but also a potential convenient prototype supersolid, allowing one to explore aspects of this intriguing phase difficult to access experimentally in other systems, such as helium. This suggests that a system of soft-core bosons may warrant further investigations, aimed at exploring additional aspects of the supersolid phase, which could render its identification in other condensed matter systems. For instance, the study of the excitation spectrum of the supersolid phase, with particular attention to the presence of two separate modes, corresponding to the simultaneous breaking of translational and gauge symmetries, seems a worthwhile undertaking.
\\ \indent
Also of interest would be the study of supersolid mixtures, which may display intriguing de-mixing properties, as well as the motion of impurities through a supersolid.
\\ \noindent
\begin{acknowledgements}
This work was supported by the Natural Science and Engineering Research Council of Canada, under Research Grant No. 121210893. The author wishes to acknowledge useful discussions with F. Cinti, P. Jain,  S. Moroni and  N. V. Prokof'ev.
\end{acknowledgements}


\begin{thebibliography}{99}
\bibitem{note}
Alternatively, one could search for  supersolid phases in systems wherein ``elementary" constituents are composite particles, e.g., Cooper pairs in superconducting Josephson junction arrays, or excitons in electron-hole bilayers. In this case,  the interaction between two such objects is of the effective kind, i.e., induced by the medium in which these particles are embedded, and has an important time-dependent  component.
\bibitem{note2} It should be noted that in this article we only consider many-body systems in {\it continuous space}, i.e., we are leaving out of the discussion supersolid phases on discrete lattices.
\end{thebibliography}
\end{document}